\documentclass[fleqn,usenatbib]{mnras}
\usepackage{comment}
\usepackage[parfill]{parskip}
\usepackage[british]{babel}             
\usepackage{newtxtext}                  
\usepackage[slantedGreek]{newtxmath}    
 \usepackage[]{natbib}  
\let\la=\lesssim 
\usepackage[T1]{fontenc}                
\usepackage{graphicx}	
\usepackage{amsmath}	
\usepackage{longtable}
\usepackage{ragged2e}
\usepackage{url}

\title[Swift J1357.2$-$0933 as seen with \emph{Swift} and \emph{NICER}] 
{Millihertz X-ray variability during the 2019 outburst of black hole candidate Swift~J1357.2$-$0933}

\author[Aru Beri  et.al. ]{Aru Beri,$^{1,2}$ \thanks{a.beri@soton.ac.uk}
Vishal~Gaur$^1$,
Phil Charles,$^2$
David R.A. Williams,$^3$
Jahanvi,$^4$
John A. Paice,$^{2,3}$
Poshak~Gandhi,$^2$
\newauthor
Diego~Altamirano,$^2$
Rob~Fender,$^{5,6}$
David~A.~Green,$^7$
David~Titterington$^7$
\\
$^{1}$Indian Institute of Science Education and Research (IISER) Mohali,~Punjab 140306, India \\
$^{2}$Physics \& Astronomy, University of Southampton, Southampton, Hampshire SO17 1BJ, UK \\
$^{3}$Jodrell Bank Centre for Astrophysics, School of Physics and Astronomy, The University of Manchester, Manchester, M13 9PL, UK \\
$^{4}$GAPHE, University of Liege, Liege 4000, Belgium \\
$^{5}$Astrophysics, Department of Physics, University of Oxford, Keble Road, Oxford OX1 3RH, UK \\
$^{6}$Department of Astronomy, University of Cape Town, Private Bag X3, Rondebosch 7701, South Africa \\
$^{7}$Astrophysics Group, Cavendish Laboratory, 19 J.J. Thomson Avenue, Cambridge, CB3 0HE, UK \\
}

\hypersetup{draft}
\begin{document}
%
%
%
\maketitle
\label{firstpage}
\begin{abstract}
Swift J1357.2$-$0933 is a black-hole candidate X-ray transient, which underwent its third outburst in 2019, during which several multi-wavelength observations were carried out.~Here, we report results from the \emph{Neil Gehrels Swift} and \emph{NICER} observatories and radio data from \emph{AMI}.~For the first time,~millihertz quasi-periodic X-ray oscillations with frequencies varying between ${\sim}$~1--5~$\rm{mHz}$  were found in \emph{NICER} observations and a similar feature was also detected in one \emph{Swift}--\textsc{XRT} dataset.~Our spectral analysis indicate that the maximum value of the measured X-ray flux is much lower compared to the peak values observed during the 2011 and 2017 outbursts.~This value is ${\sim}$~100 times lower than found with \emph{MAXI} on MJD~58558 much ($\sim$~68 days) earlier in the outburst, suggesting that the \emph{Swift} and \emph{NICER} fluxes belong to the declining phase of the 2019 outburst.~An additional soft component was detected in the \textsc{XRT} observation with the highest flux level, but at a relatively low  $L_{\rm X}$~$\sim$~$3{\times}10^{34}~(d/{\rm 6~kpc)}^2\rm{erg}~\rm{s}^{-1}$, and which we fitted with a disc component at a temperature of $\sim 0.17$~keV.~The optical/UV magnitudes obtained from \emph{Swift}--\textsc{UVOT} showed a correlation with X-ray observations, indicating X-ray reprocessing to be the plausible origin of the optical and UV emission.~However, the source was not significantly detected in the radio band.~There are currently a number of models that could explain this millihertz-frequency X-ray variability; not least of which involves an X-ray component to the curious dips that, so far, have only been observed in the optical. \\

\end{abstract}


%
%
%
%
\section{Introduction}
\label{sec:intro} 
Low mass X-ray binaries~(LMXB) are composed of a 
black-hole (BH) or neutron star (NS) accreting from a companion star of mass less than a solar mass  \citep[$M_{\odot}$, see e.g.][]{Lewin06}.
These include a significant number of transient systems that show an abrupt increase in X-ray luminosity of several orders of magnitude over a few days
followed by {\bf{a}} decay on a timescale of a few weeks, a month, or
several months \citep[e.g.][]{Frank87}.  

Swift~J1357.2$-$0933 (hereafter, J1357) is a transient LMXB discovered by \emph{Swift}~Burst Alert Telescope~(\textsc{BAT}) during its 2011 outburst \citep{Krimm11a}.~It is believed to host a black-hole
of mass $\geq9M_{\odot}$ \citep{Corral2016} in a close binary orbit with an orbital period~($P_{\rm{orb}}$) of $\rm{2.8~h}$ \citep{Corral13}, although this has not yet been confirmed.~Soon after its discovery, \citet{Rau11} reported a detection of the optical  counterpart~(at $r'=16.30$) using \textsc{GROND} images, while \textsc{SDSS} archive images revealed the presence of a pre-outburst counterpart with $r'= 21.96$.
Subsequent time-resolved photometry of J1357 in quiescence (\citealt{Shahbaz13} and \citealt{Russell17}) showed substantial variability, but no periodicity that might be associated with eclipses, dips or ellipsoidal modulation due to the donor.~Consequently, even in quiescence, the donor is not the dominant emitter at optical/IR wavelengths, making it difficult to constrain J1357's distance.  

Time-resolved optical spectroscopy \citep{Torres15} revealed the presence of prominent broad double-peaked $H{\alpha}$ emission with no hint of any late-type spectral features, such as TiO bands.
\citet{Russell17} found substantial variability in the six-year optical/IR (OIR)~light curve of J1357 during quiescence, indicating a substantial and continuing disc-emitting component.~This study suggested that J1357 lies at a greater distance, as was also discussed by \citet{Shahbaz13} who used the outburst amplitude-$P_{\rm{orb}}$ relation, extending the possible distance out to $\ge$~$6.3~\rm{kpc}$.~More recently,~\citet{Charles2019} argued similarly, thus increasing earlier luminosity estimates by ${\sim}{\times}$40, and making J1357 a member of the 
luminous, Galactic X-ray binaries instead of `very faint X-ray transients' (or VFXTs). \\
J1357 has proven to be a highly enigmatic source and several observed properties of J1357 cannot yet be explained by `standard' LMXB models.~Based on the presence of
the optical dipping whose period evolves during the outburst, \citet{Corral13} proposed that there exists a warped disc, including a thick inner torus, which is seen at a high inclination $i$ ${>}70^{\circ}$.~However,~such a geometry has difficulties explaining the lack of eclipses by the donor or any X-ray dipping, as well as the lack of X-ray reflection features \citep[see][for details]{Beri19,Paice19}.~An updated multi-component model of J1357 
to account for these properties has been proposed by \citet{Paice19}, which requires a truncated accretion disc, with an inner disc radius much greater than the radius of the innermost stable circular orbit~($R_{\rm{ISCO}}$).
They suggest that the presence of an extended X-ray corona in between the black-hole and  accretion disc could explain the lack of X-ray reflection features and absence of X-ray dips. They also propose a jet region (synchrotron emission) near the black-hole observed at a high inclination and sporadically occluded by the accretion disc's vertical extensions, causing significant dips in the red band optical light curves.~However, these authors also noted that this model could not explain all the source properties, such as  why the perturbations in the accretion disc move outwards during outburst. 

In X-rays there has only been one detection of a quasi-periodic oscillation (QPO) from J1357, and that was a low-frequency~($\sim$~6~\rm{mHz}) QPO seen at the start of the 2011 outburst \citep{Armas13b} by the Rossi X-ray Timing Explorer~(\emph{RXTE}).~During the 2017 outburst of J1357, high-speed multi-wavelength photometry was performed, revealing the lack of any X-ray response to the peculiar optical dips.~The lack of X-ray dips in the light-curves were explained as being due to the presence of a truncated disc with an extended X-ray corona.~However, the detailed accretion geometry remains poorly understood, and so further high time-resolution X-ray studies are clearly required. 

Such an opportunity presented itself early in 2019 when J1357 underwent its third outburst \citep{Velzen2019,Gandhi2019,Russell2019,Beri2019b}.~Here we report on observations with the \emph{Neil Gehrels Swift} observatory,~X-ray data from the Neutron star Interior Composition Explorer~(\emph{NICER}) and radio data from the Arcminute Microkelvin Imager~(\emph{AMI}).~ In \S \ref{sec:dataanalysis}, we give observational and data reduction details, while spectral results are presented in \S \ref{sec:spectra}.~Timing analysis and results are presented in \S \ref{Timing}.~\S \ref{sec:multi-band} gives details on results from UV/Optical and radio observations.~We discuss our new results in \S \ref{Discuss}. 

\begin{table*}
\caption{Log of the \emph{Swift} observations and \textsc{XRT} spectral results during the 2019 outburst of J1357. }

\begin{tabular}{ c c c c c c c c c c}
\hline

Obs-ID & Time & Mode & Count rate  & Exp-time~ & $\Gamma$ & $F_{\rm X,unabs}$ &  $L_{\rm X}^{\dagger}$  & cstat/dof \\

       & $\rm{MJD}$    &         & $\rm{count~s^{-1}}$ & $\rm{ks}$ &     &  $10^{-12} \rm erg~\rm cm^{-2}~\rm s^{-1}$ & $10^{32} \rm erg~\rm s^{-1}$ & \\
\hline
$31918084^{*}$ & $58630.2533$ & WT & $0.22\pm0.02$ & 1.3 & $2.3\pm0.3$   & $8.2\pm{1.0}$     & $347\pm42$ & 51/64 \\
$31918085$ & $58632.2957$ & WT & $0.11\pm0.01$ & 1.5 & $2.0\pm0.2$   & $4.9\pm{0.4}$    & $212\pm19$  & 160/155 \\
$88872001$ & $58633.0559$ & PC & $0.16\pm0.01$  & 1.1 & $1.7\pm0.1$   & $6.6\pm{0.4}$    &   $284\pm34$ & 97/133  \\
$31918086$ & $58636.5232$ & PC & $0.15\pm0.01$ & 1.6 & $1.6\pm0.2$   & $6.0\pm{0.3}$       & $258\pm25$ & 144/166 \\
$31918087$ & $58642.2938$ & PC & $0.14\pm0.01$ & 1.6 & $1.8\pm0.1$   & $4.8\pm0.3$         &  $207\pm12$ & 133/144\\ 
$31918088$ & $58649.1678$ & PC & $0.11\pm0.01$ & 1.4 & $1.6\pm0.1$   & $2.5\pm0.3$     &  $105\pm12$ &  89/110  \\
$31918089$ & $58654.1139$ & PC & $0.08\pm0.01$ & 0.55 & $1.6\pm0.2$   & $2.4\pm0.2$       &  $107\pm12$ & 38/41    \\
$31918090$ & $58660.1249$ & PC & $0.10\pm0.0086$ & 1.6 & $1.9\pm0.2$   & $3.0\pm0.2$   & $129\pm12$ & 87/97   \\
$31918091$ & $58662.6706$ & PC & $0.06\pm0.01$ & 0.7  & $1.6\pm0.2$  & $3.8\pm0.6$     & $161\pm12$  &19/38    \\
$31918092$ & $58668.5556$ & PC & $0.05\pm0.01$ & 0.6 & $2.4\pm0.5$   & $0.7\pm0.2$      & $29\pm7$ & 10/21     \\
$31918093$ & $58672.5342$ & PC & $0.07\pm0.01$ & 1.0 & $1.6\pm0.3$   & $1.9\pm0.5$     &  $81\pm12$ &         28/66 \\
$31918094$ & $58682.1041$ & PC & $0.06\pm0.006$  & 1.4  & $2.1\pm0.2$  & $1.7\pm0.2$     &  $73\pm12$  &  43/70      \\
$31918095$ & $58692.9878$ & PC & $0.008\pm{0.003}$ & 1.3  & $1.2\pm{1.0}$  &  $0.2\pm0.1$     & $9\pm2$ & 4/7    \\
$31918096$ & $58703.8096$ & PC & $0.006\pm{0.002}$ & 1.5  & $1.4\pm0.9$  & $0.2\pm0.1$      & $9\pm2$ &  5/8 \\

$31918097$ & $58712.3114$ & PC & ${0.0035}^{c}$ & 1.3  & $2.5$ & $<{0.102}^a$      &  $<{4.4}^{a}$ & - \\ 

\hline

\end{tabular}
\label{Swift}
\\
{\begin{flushleft}
Notes: \\
$^*$ Shows a sharp mHz peak in the PDS at a 95{\%} significance level. \\
$^{\dagger}$ Calculated from the 0.5–10 keV unabsorbed flux~($F_{\rm X,unabs}$) for a distance of $6.0~\rm{kpc}$ \\
$^a$~ 95$\%$ confidence upper level limit count rates using the prescription given by \citet{Gehrels86}.~We estimated the corresponding unabsorbed flux upper limits using the \textsc{WEBPIMMS HEASARC} tool.  \\

\end{flushleft}
}
\end{table*}

 \begin{table*}
 \def\arraystretch{1.2}%
\caption{Spectral parameters of J1357 for an absorbed disk-blackbody and power-law model from \emph{Swift}-\textsc{XRT} data.~We have assumed a distance of $6.0~\rm{kpc}$ for estimating the diskbb radius. }
\label{XRT-bbodyrad} 
\begin{tabular}{cccccc}
 \hline
 Obs~ID &  kT & Diskbb Radius & $\Gamma$ & Power-law norm  & cstat/dof  \\  

        &  \rm{keV}  &  \rm{km} &       &   &   \\
\hline
31918084  &   $0.17\pm0.06$     &   $25_{-17}^{+45}$           &   $0.4\pm0.6$       &  ($3.1_{-1}^{+3}$)${\times} 10^{-4}$ & 31/62               \\   
\hline
\end{tabular}
\end{table*}

\begin{table*}
\caption{Log of the \emph{NICER} observations during the 2019 outburst of J1357.}
\label{NICER}
\begin{tabular}{ c c c c c c c c c c c c}
\hline

Obs-ID & Time~ &  Exp-time & Count rate & $\Gamma$ & $F_{X, unabs}$   & $L_{X}^{\dagger}$ & $\chi^{2}$/dof\\

       & $\rm{MJD}$        & $\rm{ks}$             & $\rm{count~s^{-1}}$     &      &  $10^{-12} \rm erg~\rm cm^{-2}~\rm s^{-1}$ & $10^{32} \rm erg~\rm s^{-1}$ \\
\hline

2200730101~(Obs~1) &  58626.57  &  1.9 & $3.47\pm0.05$ & $1.86\pm0.03$ &  $7.25\pm0.10$ & $312.3\pm4.2$ & 143/130\\
2200730102~(Obs~2) &  58627.02  &  1.8 & $3.65\pm0.05$ & $1.86\pm0.03$ & $7.49\pm0.11$ & $322.4\pm4.8$ & 116/124\\
2200730103~(Obs~3) &  58629.41 &  6.4 & $3.30\pm0.03$ & $1.84\pm0.02$ & $7.28\pm0.05$ &  $313.7\pm2.3$ & 157.8/157 \\
2200730104~(Obs~4) &  58629.99 &  1.5 & $3.63\pm0.05$ & $1.80\pm0.03$ & $7.89\pm0.12$ &  $339.6\pm5.0$ & 118/119 \\
2200730105~(Obs~5) &  58635.53 &  2.8 & $2.65\pm0.04$ & $1.85\pm0.03$ & $5.84\pm	0.07$ & $251.3\pm3.1$ & 132/141 \\
2200730106~(Obs~6) &  58636.05 &  3.3 & $2.66\pm0.03$ & $1.87\pm0.03$ & $5.82\pm0.07$ &  $250.7\pm2.9$ & 123/147 \\ 
2200730107~(Obs~7) &  58637.01 &  5.4 & $2.60\pm0.03$ & $1.88\pm0.02$ & $5.75\pm0.05$ &  $247.5\pm2.3$ & 151/155 \\
2200730108~(Obs~8) &  58639.98 &  1.4 & $2.15\pm0.04$ & $1.87\pm0.05$ & $5.12\pm0.11$  & $220.5\pm4.7$ & 789/810 \\
2200730109~(Obs~9) &  58641.91 & 2.2 & $2.15\pm0.04$ & $1.92\pm0.04$ & $4.18\pm0.08$ &  $179.9\pm3.3$ & 105/107 \\
2200730110~(Obs~10) &  58645.90 &  1.3  & $1.89\pm0.05$ & $1.96\pm0.05$ & $3.79\pm0.08$ &  $163.4\pm3.5$ & 862/910\\
2200730111~(Obs~11) &  58648.42 &  2.8 & $2.19\pm0.03$ & $1.90\pm0.03$ & $4.75\pm0.06$ &  $204.4\pm2.8$ & 151/128 \\ 2200730112~(Obs~12) &  58653.32 &  0.84 & $1.87\pm0.06$ & $1.87\pm0.06$ & $4.66\pm0.11$ &  $200.7\pm4.9$ & 650/650 \\
2200730113~(Obs~13) &  58654.09 &  2.5  & $1.65\pm0.03$ & $1.88\pm0.04$ & $4.05\pm0.06$ &  $174.3\pm2.8$ & 106/113 \\
2200730114~(Obs~14) &  58655.64 &  0.6  & $1.31\pm0.06$ & $1.99\pm0.08$ & $3.00\pm0.11$ &  $129.1\pm4.6$ & 417/350 \\
2200730115~(Obs~15) &  58656.35 &  3.3  & $1.44\pm0.03$ & $1.95\pm0.04$ & $3.25\pm0.05$ &  $139.8\pm2.2$ & 141/120 \\
2200730116~(Obs~16) &  58656.99 &  3.2  & $1.68\pm0.0.03$ & $2.01\pm0.04$ & $3.28\pm0.06$ & $141.3\pm2.4$  & 92/106 \\
2200730117~(Obs~17) &  58663.38 &  2.4  & $1.50\pm0.03$ & $1.98\pm0.05$ & $3.01\pm0.06$ &  $129.6\pm2.6$ & 87/100 \\
2200730118~(Obs~18) &  58664.22 &  3.0  & $1.48\pm0.03$ & $2.06\pm0.04$ & $2.81\pm0.05$  & $121.0\pm2.1$ & 102/117 \\
2200730119~(Obs~19) &  58667.06 &  0.25 & $0.97\pm0.08$ & $2.14\pm0.19$ & $1.79\pm0.13$ &  $77.2\pm5.8$ & 9.2/11 \\
2200730120~(Obs~20) &  58682.46  &  2.5 & $0.80\pm0.02$ & $2.12\pm0.06$ & $1.31\pm0.03$ &  $56.5\pm 1.4$ & 78/73 \\
2200730121~(Obs~21) &  58688.01  &  3.5 & $0.06\pm0.01$ & $2.23\pm0.10$ & $0.48\pm0.02$ &  $20.5\pm 0.9$ & 72/68 \\

\hline
\end{tabular}
{\begin{flushleft}
Notes: \\
$^{\dagger}$ Calculated from the 0.5–10 keV unabsorbed flux~($F_{\rm X,unabs}$) for a distance of $6.0~\rm{kpc}$ \\
\end{flushleft}
}
\end{table*}

\begin{figure}
\includegraphics[height=1.5\columnwidth,width=2\columnwidth,angle=0, keepaspectratio]{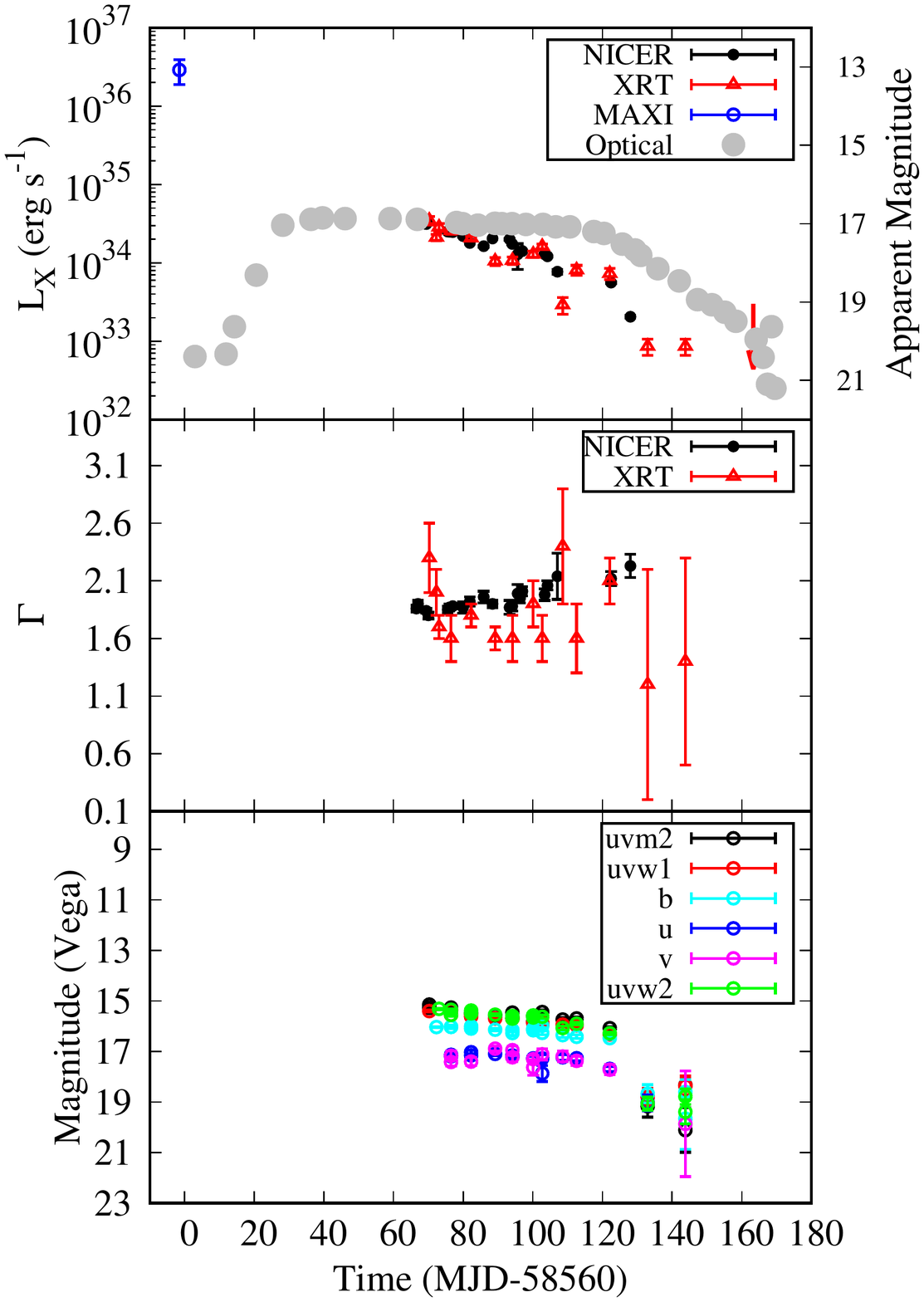}
\caption{Overview of the J1357 2019 outburst. The {\it (upper panel)} shows $L_{X}$ (for $d=6$~\rm{kpc}) using data from \emph{Swift}--\textsc{XRT} and \emph{NICER} (0.5--10~\rm{keV}) and an outline (grey discs) of the ground-based $r$-band monitoring from \citet{Russell2019}.~The {\it (middle panel)} shows the evolution of photon index, {$\Gamma$}, and the {\it (bottom panel)} shows UV/optical magnitudes (Vega system). Time is in days relative to MJD 58560, the estimated start date of the outburst.}
\label{PL-index}
\end{figure}

\begin{figure}
\includegraphics[height=1\columnwidth,width=\columnwidth,angle=0,keepaspectratio]{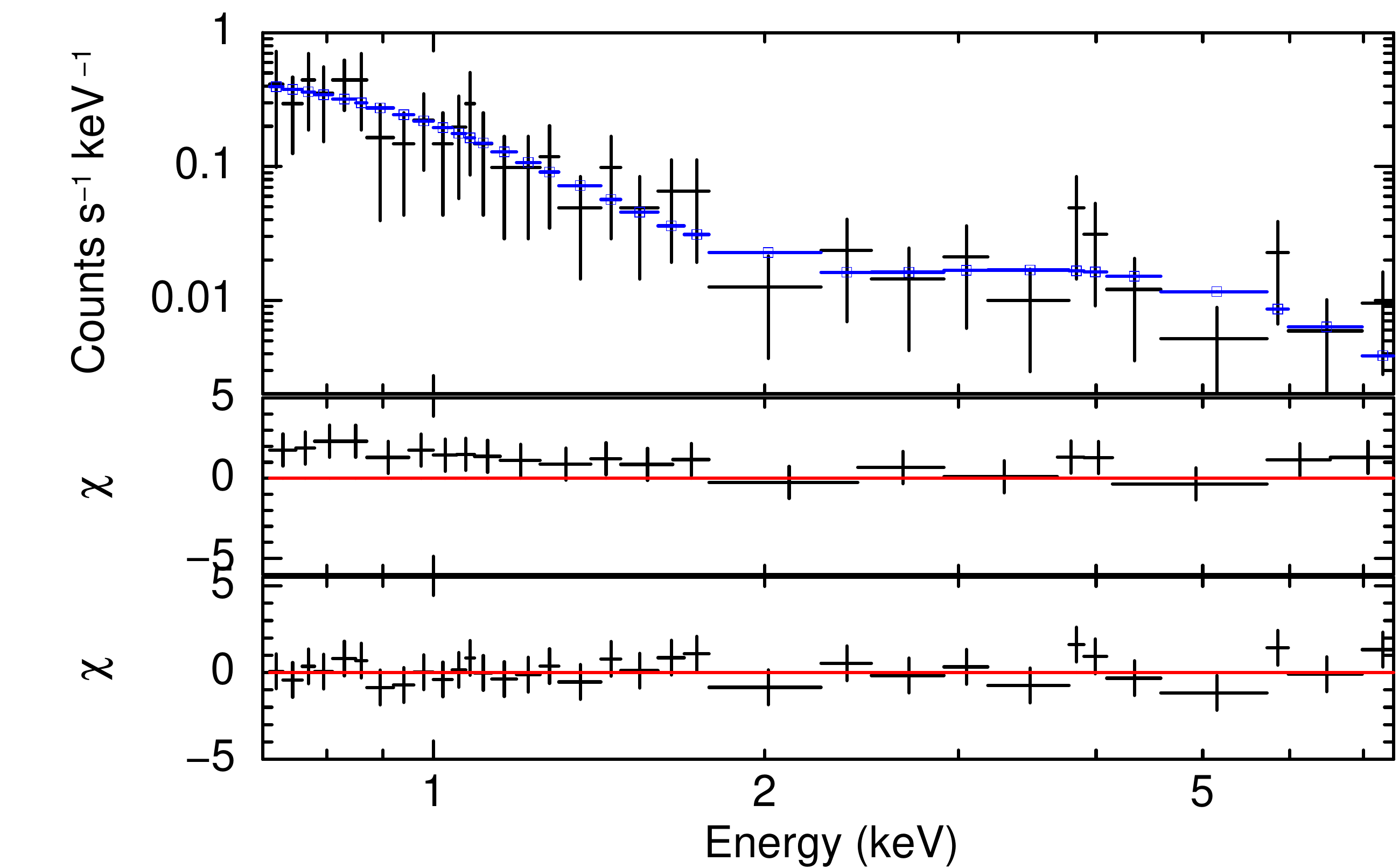}
\caption{{\it (Top panel)} Best-fitting \emph{Swift}--\textsc{XRT} spectra of J1357, obtained using an absorbed diskbb and power-law.~The lower panels show residuals ({$\chi$}) to the fits of ({\it middle panel}) a simple absorbed power-law (which reveals a low-energy excess), and ({\it bottom panel}) the best-fit model.} 
\label{Spectra}
\end{figure}
\section{Observations and data analysis}
\label{sec:dataanalysis}
\subsection{\emph{Swift}}
We found that weak X-ray activity was detected in \emph{MAXI} around 
MJD 58558 (see Figure~\ref{PL-index}),~and estimated the corresponding unabsorbed flux value using the WEBPIMMS HEASARC tool which translates to $L_X$ ~${\sim}3{\times}10^{36}{\rm{erg~s^{-1}}}$ for $d=6~\rm{kpc}$.
This was also noted by \citet{Gandhi2019} and \citet{ Russell2019}, and these authors suggested that the 2019 outburst started around MJD~58562. 

We have used data from the X-ray Telescope~(\textsc{XRT}) and the Ultraviolet and Optical Telescope~(\textsc{UVOT}) instruments on-board the~\emph{Neil Gehrels Swift} Observatory~\citep{Gehrels04}.~The \textsc{XRT}
is sensitive in the 0.2--10~\rm{keV} energy range and has an effective area of $100~\rm{cm^2}$ \citep[see][for details]{Burrows05}
while \textsc{UVOT} covers UV and optical bands~(170--600~\rm{nm}) \citep[][]{Roming04}. 

We obtained a total of 15 observations of J1357 between May 27 and August 8,~2019, totalling $\sim$18~\rm{ks} of exposure time (see Table~\ref{Swift} for details).
The first two \textsc{XRT} observations were in the windowed timing (WT) mode
while the rest were in photon counting (PC) mode.
The online tools provided by the UK Swift Science
Data Centre\footnote{http://www.swift.ac.uk/} \citep{Evans09}
were used to obtain light curves and spectra from these data. 

\textsc{UVOT} observations were taken in image mode, the majority using the six available filters $(v, b, u, uvw1, uvm2, uvw2)$.
 Light curves in each filter were created using the \textsc{uvotmaghist} tool that uses \textsc{uvotsource} to perform aperture photometry on all sky images (in each filter) available for an individual observation.~Source magnitudes were computed in the Vega system using a 5~arcsec radius circle centred on J1357, together with a neighbouring source-free 10~arcsec radius circle for background correction.~All flux values and magnitudes were corrected for Galactic extinction in this direction ($E(B-V)=0.04$) as done by 
 \citet{Armas13a} and \citet{Beri19}. 
 
\subsection{\emph{NICER}}
 The \emph{NICER} X-ray Timing Instrument \citep{Gendreau2016} has 56 concentrators, each coupled to a silicon drift detector housed in a Focal Plane Module~(FPM).~At the time of J1357 observations, 52 of the 56 FPMs were functional, providing an effective area of ${\sim}1750~\rm{cm^2}$ in the 0.2--12~${\rm{keV}}$~band.
 J1357 was observed with \emph{NICER} between May 23 and August 20,~2019 (see Table~\ref{NICER}), and the data were reprocessed using \textsc{NICERDAS} version 8c,
 distributed as part of \textsc{HEAsoft version~6.29c}.~The cleaned event files were generated using the  \textit{nicerl2}\footnote{https://heasarc.gsfc.nasa.gov/lheasoft/ftools/headas/nicerl2.html} task in \textsc{HEAsoft version~6.29c} using the default filter criteria which includes limiting analysis to time intervals with a pointing offset $<54$~arcsec, a bright Earth limb angle $>$30$^{\circ}$, a dark Earth limb angle $<$15$^{\circ}$, and ignoring data collected during the passage through the South Atlantic Anomaly.~Standard screening procedures also include rejecting all time intervals where the rate of saturating particle events~(overshoots)~is greater than 1~$\rm{counts~s^{-1}detector^{-1}}$.~These cleaned events were then filtered using the {`\it{nifpmsel}'} tool to remove data from two noisier detectors~(Detectors~14 and 34).~The final cleaned events (from all MPU) were then processed using \textsc{xselect} to obtain scientific products.~The background spectra were generated using the {`\it{nibackgen3C50}'} tool \citep{Remillard2021}.~Standard tools ({`\it{nicerrmf}'} and {`\it{nicerarf}'}) were used to create response and auxiliary files for the spectral analysis.~In order to investigate the presence of any artificial features such as dips in the X-ray light curves, we also examined data following the procedure recommended by the \emph{NICER} team\footnote{https://heasarc.gsfc.nasa.gov/docs/nicer/analysis\_threads/iss\_obstruction/}. 
 
 \subsection{\emph{Arcminute Microkelvin Imager~(AMI)}}
 Radio observations with the \emph{Arcminute Microkelvin Imager Large Array} ~\citep[AMI-LA,][hereafter AMI]{AMILA, AMILA2} were triggered after the source went into outburst. AMI is an 8-element radio array located in Cambridge, UK, operating at a central frequency of 15.5\,\rm{GHz} with a 5\,\rm{GHz} bandwidth. A single observation of J1357 was performed on 2019 May 26~(MJD~58629.84) for a total of four hours. The observation used six of the eight 13~m AMI antennas, interleaving 100~s scans of the phase calibrator NVSS J140412$-$001324 with 10 minute scans of the target field. The data were flux calibrated using 3C286 and were reduced with a custom pipeline \textsc{REDUCE$\_$DC} \citep{Perrott2015} and imaged using the Common Astronomical Software Applications~(\textsc{CASA})\footnote{\url{https://casa.nrao.edu/}}. Unfortunately, NVSS~J135720$-$093003 is $<$3~arcmin away from J1357 and within the primary beam of \emph{AMI}. It was removed from the \textit{uv}-data using the \textsc{CASA} task \texttt{uvsub} and a final image was obtained of the field, with an rms-noise limit of 70~$\upmu$Jy~beam$^{-1}$.~J1357 is not detected above a 3$\sigma$ rms noise~(210~$\upmu$Jy~beam$^{-1}$) threshold. 
 \begin{figure}
\includegraphics[height=\columnwidth,width=1\columnwidth,angle=0,keepaspectratio]{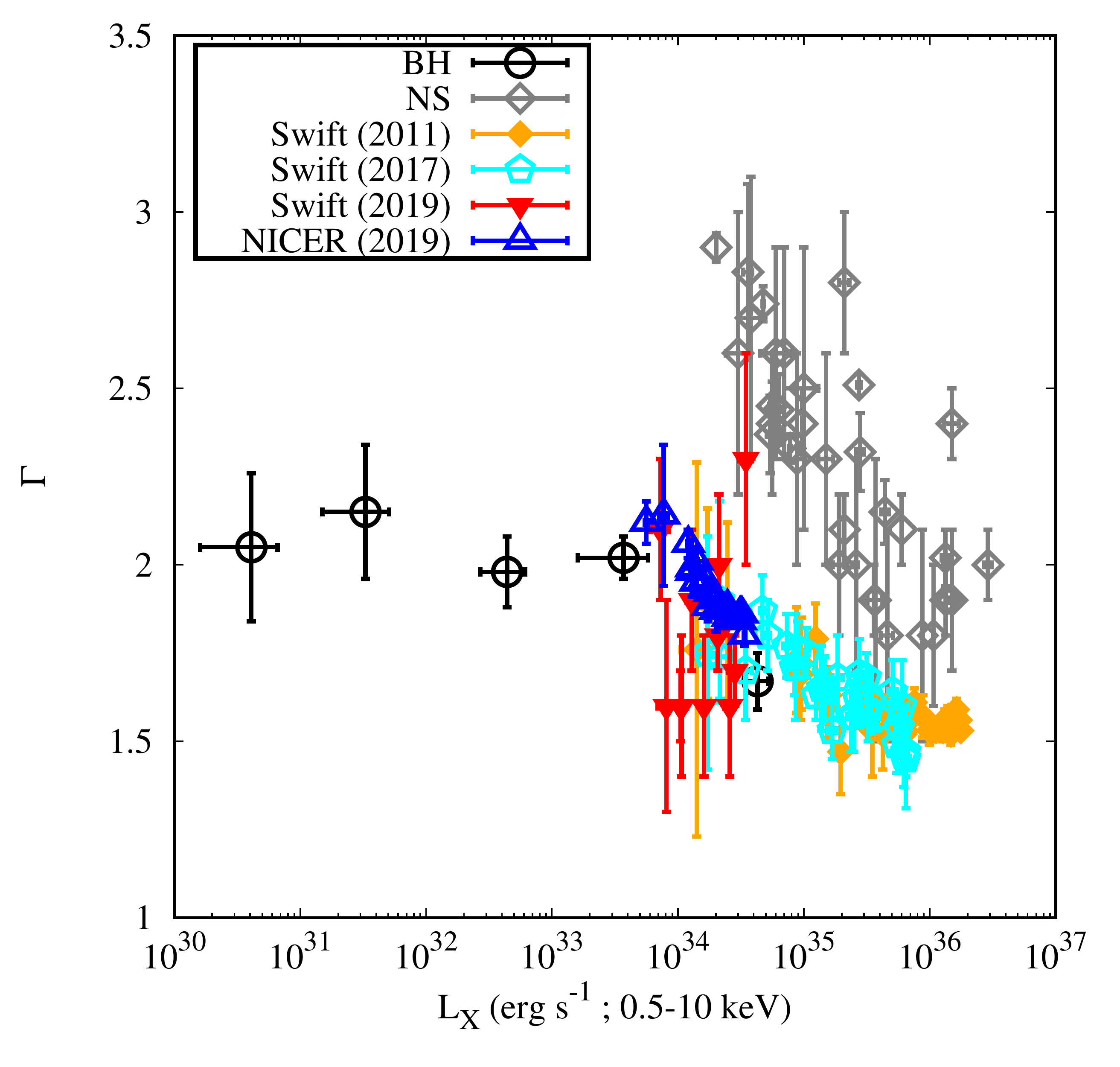}
\caption{Comparison of the spectral (photon index $\Gamma$)-- X-ray luminosity ($L_X$, 0.5--10~\rm{keV}) behaviour of J1357 (coloured points, see legend, assuming $d$ = 6~kpc) with other low-level accreting LMXBs as compiled by \citet{Wijnands15} for both BH (solid black points) and NS (grey points) transients. The earlier J1357 values are from \citet{Armas13a} and \citet{Beri19} for the 2011 and 2017 outbursts respectively. }
\label{wijnands}
\end{figure}

\section{X-ray Spectroscopy}
\label{sec:spectra}

\subsection{X-ray Spectral Analysis}
We performed spectral analysis using \textsc{xspec} 12.12.0 \citep{Arnaud96}.
Owing to the limited \emph{Swift}-\textsc{XRT} statistics, we grouped spectra using the ftools task \textit{grppha} so as to have at least one count per bin.~W-statistics (background subtracted Cash statistics) were employed while performing the X-ray spectral fitting \citep{Wachter79}.~Spectra obtained with \emph{NICER} were grouped into bins with a minimum of 25 photons and $\chi^2$ statistics were applied.
Interstellar absorption was included in all our spectral fits, employing \citet{Wilms00} abundances and \citet{Verner96} photo-electric cross-sections.
\texttt{TBABS} was used to model the hydrogen column density~($N_{\rm{H}}$) with a value fixed to $1.2 \times 10^{20}~\rm{cm}^{-2}$ as obtained by \citet{Armas13b} from the \emph{XMM-Newton} high-resolution X-ray
spectra.
The X-ray spectral fitting was performed over ranges of 0.5--10~$\rm{keV}$ for PC mode and 0.7--10~$\rm{keV}$ for WT mode, due to the presence of low energy residuals in the latter\footnote{ http://www.swift.ac.uk/analysis/xrt/}.~For the \emph{NICER} spectra, emission from the source was detectable over a narrow energy range in all cases (0.4–2.5~\rm{keV}), and so our spectral fits are confined to this range.
All the fluxes reported are unabsorbed 0.5--10~$\rm{keV}$ fluxes obtained using the convolution model \textsc{`cflux'}.~Unless explicitly mentioned, we quote
all errors at 1-${\sigma}$ confidence level. 

\subsection{X-ray Spectral Results}
The top panel of Figure~\ref{PL-index} shows the $L_{\rm X}$~(0.5--10~\rm{keV}) from \emph{Swift}-\textsc{XRT} and \emph{NICER}.
The maximum value of unabsorbed flux measured is ${\sim} 8\times 10^{-12}~\rm{erg}~\rm{cm}^{-2}~\rm{s}^{-1}$ which corresponds to $L_{\rm X}$ of about 
$3.4{\times}10^{34}~($d$/6~{\rm kpc})^2{\rm erg}~{\rm s^{-1}}$,
lower than previous outbursts, likely as a result of not observing until well after the outburst had started and declines throughout the \emph {Swift} and \emph{NICER} observations~(Table~\ref{Swift} \& \ref{NICER}).~We have also replotted the published optical outburst profile from \citet{Russell2019}.~Using an absorbed power-law model, we find that
the power-law index~($\Gamma$) evolves over the outburst~(Figure~\ref{PL-index} middle panel) with $\Gamma$ between $\sim$1.5 and $\sim$2.5 during these observations, indicating the presence of soft X-ray spectra. 

In one of the \textsc{XRT} spectra~(ObsID~31918084) with higher statistics,~we found that using an absorbed power-law model showed some excess in the spectral residuals~(see Figure~\ref{Spectra}).~Therefore, we added a \textit{diskbb} component to the absorbed power-law model and found significant improvement in the spectral fit, giving a disc temperature of ~0.17~keV (see Table~\ref{XRT-bbodyrad}).~To verify the statistical significance of adding a disc component, we performed an \textit{ftest} to determine the probability
of chance improvement (PCI), which was very low ($4\times10^{-7}$), indicating it is reasonable to add the diskbb component in the X-ray spectra.~However, we would like to add a caveat that given limited statistics, it was difficult to constrain all fit parameters.~To further evaluate the chance probability of improvement by adding the extra \textit{diskbb}
component, we simulated 100,000 data sets using \textit{simftest} in
\textsc{XSPEC}. The evaluated chance probability is less than $10^{-6}$, indicating that the presence of an additional disc component is significant.

The unabsorbed 0.5--10~$\rm{keV}$ flux during this \textsc{XRT} observation is ${\sim}~8{\times}10^{-12}\rm{erg~cm^{-2}~s^{-1}}$, to which the thermal component contributes 50~$\rm{per~cent}$, i.e. an inferred $L_X$~${\sim}3.4{\times}10^{34}$~$\rm{erg~s^{-1}}$.
The spectral residuals did not indicate the presence of a neutral iron $K_{\alpha}$ line at 6.4~keV, which is consistent with the previous study carried out by \citet{Beri2019b} using  \emph{Swift} and \emph{NuSTAR} data.~The non-detection of the iron line was also confirmed from our simulations of 10,000 data sets that returned the null-hypothesis probability of about 0.98. 

In Figure~\ref{wijnands}, we compare J1357's spectral properties as a function of $L_{\rm X}$
for all three outbursts,~together with those from the \citet{Wijnands15} survey of low-level accreting NS and BH transients at $L_X$ <$10^{36}~\rm{erg~s^{-1}}$.~NS LMXBs are significantly softer than BH systems below an $L_X$~($0.5-10~\rm{keV}$) of $10^{35}\rm{erg~s^{-1}}$.
During all three outbursts of J1357, the power-law index
showed similar behaviour,~clearly following
the general trend of the BH sample.~However, an interesting point to note is that the 2019 outburst showed higher values of $\Gamma$ compared to the earlier outbursts of J1357, indicating softer X-ray spectra. 



\begin{figure}
\includegraphics[height=1.\columnwidth,width=1.\columnwidth,angle=0,keepaspectratio]{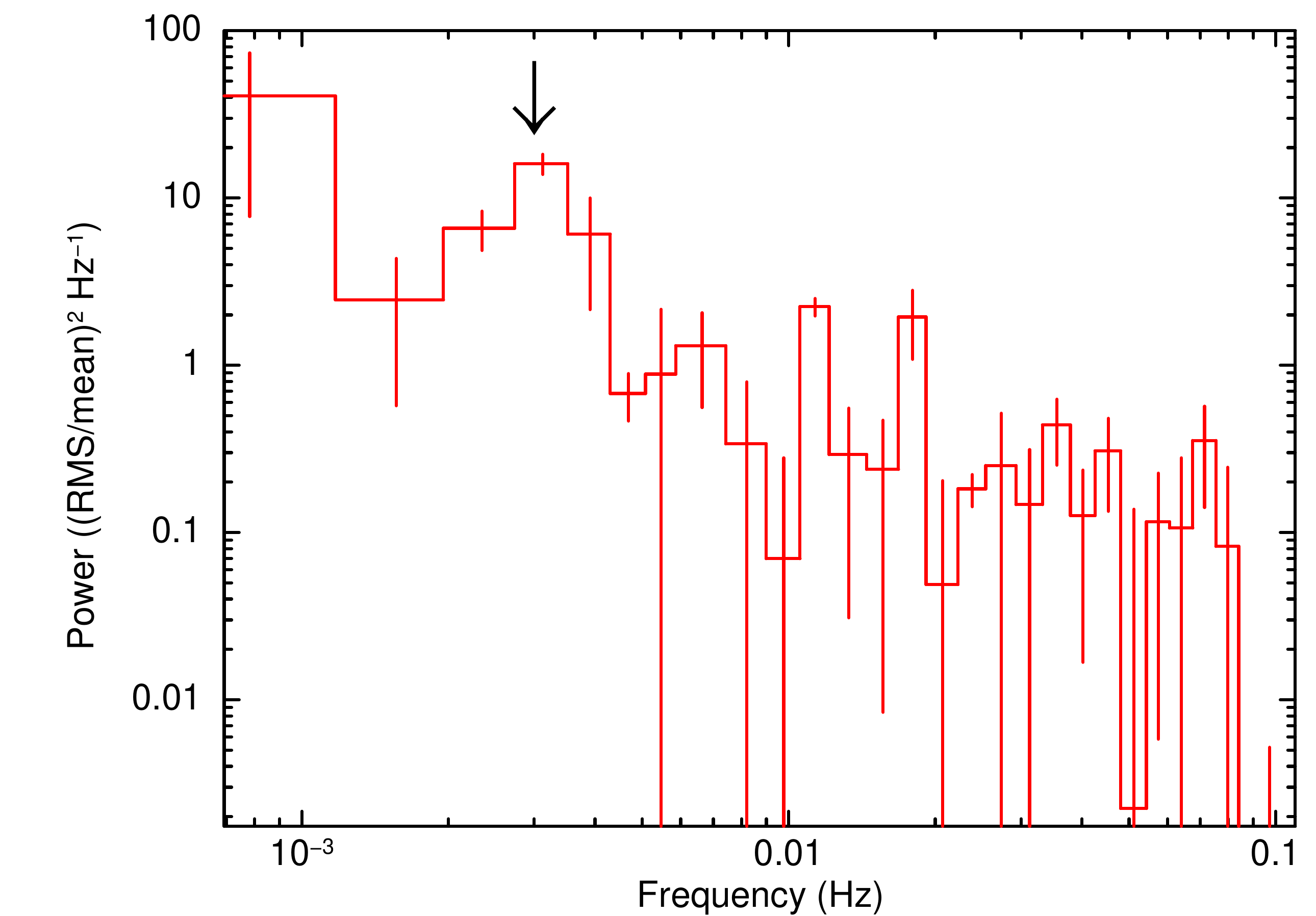}
\caption{Power density spectrum obtained using the \emph{NICER} data of Obs~9, showing the presence of a millihertz QPO (marked with an arrow) with a rms fractional variability of${\sim}~4$ per cent ($0.2-12~\rm{keV}$).}
\label{PDS}
\end{figure}
\begin{figure*}
\centering
\includegraphics[height=8\columnwidth,width=2\columnwidth,angle=0,keepaspectratio]{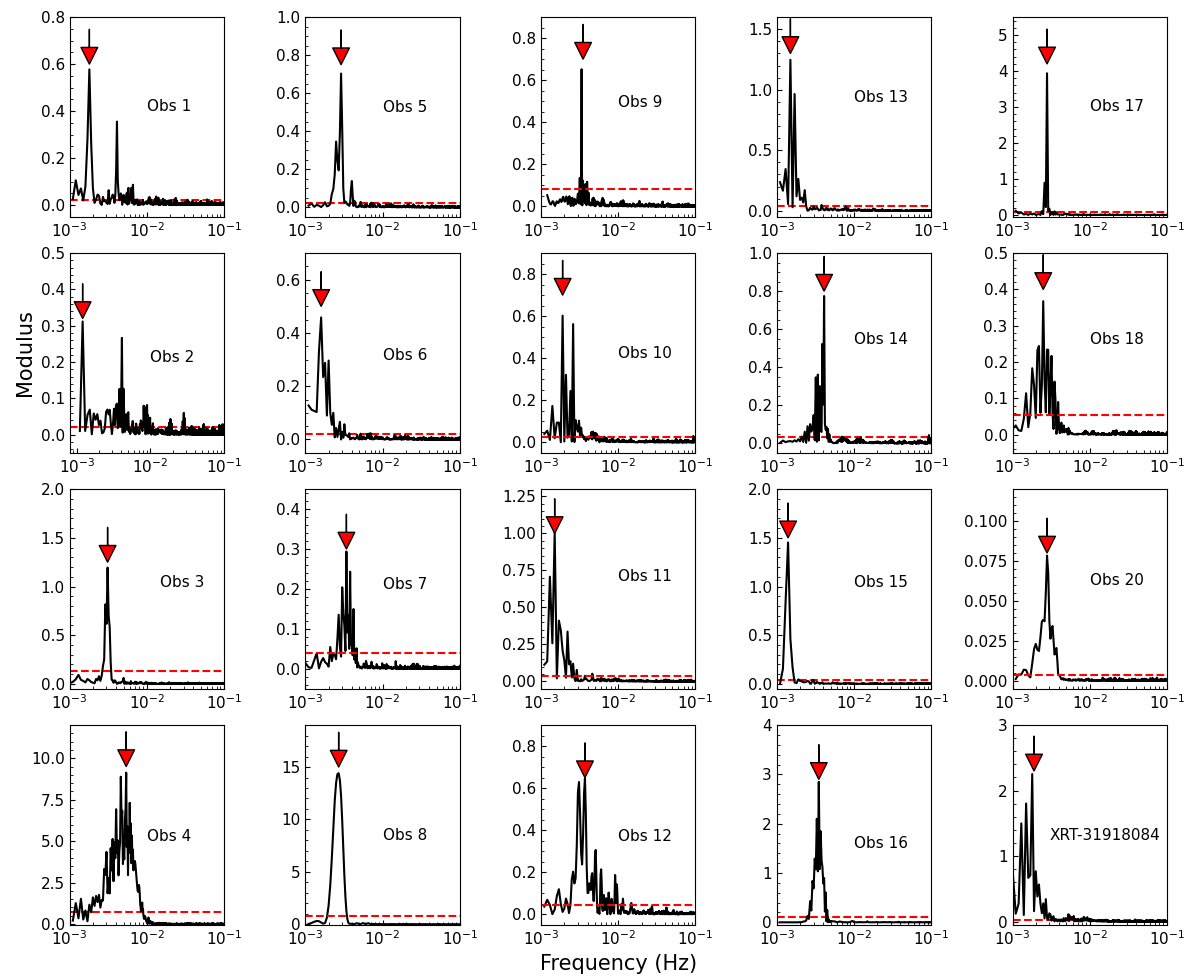}
\caption{The plots show CLEAN power density spectra. The arrows indicating sharp peaks in the millihertz frequency range.~We have included only those observations during which the false alarm probabilities for the period detection lie below 0.01 with 95 per cent confidence.~The dashed line (red) is the $5{\sigma}$ significance level.}
\label{Periodogram}
\end{figure*}


\begin{figure}
\includegraphics[height=1.5\columnwidth,width=1.\columnwidth,angle=0,keepaspectratio]{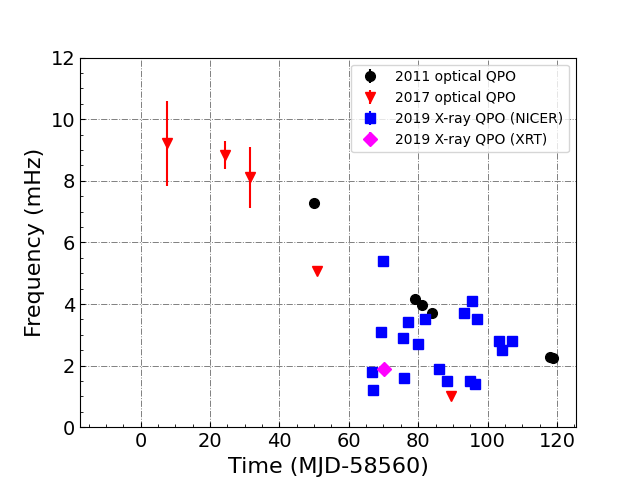}
\caption{Evolution of J1357's QPO frequency with time during the 2019 outburst (blue squares - \emph{NICER}; magenta diamond - \emph{Swift}), compared with the optical QPO behaviour in previous outbursts (see legend) from \citet{Corral13} and \citet{Paice19}.}
\label{evol}
\end{figure}
\section{Timing Results}
\label{Timing}
We investigated the X-ray light curves of J1357 to search for the presence of dip-like features similar to those observed in the optical \citep[e.g.,][]{Corral13}.~The X-ray light curves were extracted with a bin size of $100~\rm{ms}$ in the 0.2--12$~\rm{keV}$~energy range,~with average count rates for each observation given in Table~\ref{NICER}. 

We created a power density spectrum~(PDS) using data of Obs~9 with the FTOOL task `\textit{powspec}'.~Obs~9 showed the most flux variability while similar features were also seen with other \emph{NICER} observations, albeit during brief snapshots~($\sim$ 500~\rm{s}).~The Obs~9 light curve with $20~\rm{s}$ bins~
(0.2--12~$\rm{keV}$) was used to search for signals in the lower frequency range.~The PDS showed the presence of excess power around ${\sim}3~\rm{mHz}$~(Figure~\ref{PDS}), similar to that seen by \emph{RXTE} during J1357's 2011 outburst.~In order to measure the characteristic frequency~and full width at half maximum,~we fitted the Obs~9 PDS with a single, zero-centred Lorentzian, obtaining $\nu_0$ and FWHM of $0.0028\pm{0.0001}$ and $0.0007\pm{0.0003}$~\rm{Hz}.~The quality factor~($Q$) is $\sim$ $4.0\pm1.6$ and the root mean square~(rms) amplitude  is $4.8_{-1.3}^{+0.5}$ per cent.~We also observed two additional sharp peaks in the PDS around $\sim~0.011$~and 0.014~\rm{Hz}. 

In order to further investigate all these peaks in the PDS we used independent methods such as Lomb--Scargle periodogram  \citep{Lomb1976, Scargle1982, Horne1986} and CLEAN
\citep{Roberts87} as implemented in the \textsc{PERIOD} program distributed with \textsc{Starlink Software Collection\footnote{http://starlink.eao.hawaii.edu/starlink}} \citep{Currie14}.~The CLEAN algorithm is a powerful tool that basically deconvolves the spectral window from the discrete Fourier power spectrum (or dirty spectrum) and produces a CLEAN spectrum, which is largely free of the many effects of spectral leakage.
A sharp peak was observed in all the periodograms, and a consistent value of QPO frequency was obtained using this method (Figure~\ref{Periodogram}).~The QPO frequency varied between 1--5$~\rm{mHz}$ during these observations.~However, we did not observe any sharp features around 11 and 14~$\rm{mHz}$ as seen in Figure~\ref{PDS}. \\
The significance of these features was determined using a Fisher randomization test \citep{Nemec85}, which
consists of calculating the periodogram of a new, randomized time-series. This randomization of data and periodogram calculation loop is then performed for a large number of permutations.
The False Alarm Probabilities~(FAP) were calculated to indicate the significance of any peak detected, and were found to be zero~(95{\%} confidence interval 0.0 to 0.01) for those shown in Figure~\ref{Periodogram} and the values are listed in Table~\ref{peak}.~The lower significance value implies that the obtained period is correct.~A large value of FAP would suggest that any detected periodicities were unlikely to be real.
Although all X-ray light curves showed the presence of variability,~a PDS as shown in Figure~\ref{PDS} could not be created for Obs~19 with \emph{NICER} due to its short duration. 

We also evaluated the significance of the QPO against the red noise using a Monte Carlo simulation \citep[see e.g.][]{Benclloch2001, Vaughan2005}.~ This method involves determination of the shape of the PDS of each observation by fitting a power law model (as shown in Figure~\ref{Obs109-sig}).~Thereafter, we generated $10^5$ trial light curves from a best-fitting power index of PDS using Python libraries of Stingray\footnote{Stingray is a Python package for X-ray astronomy, and is available at https://github.com/StingraySoftware/stingray} Version 1.1 \citep{matteo_bachetti_2022,Huppenkothen2019b, Huppenkothen2019a}.~We have used duration, mean count rate and variance as that of the observed light curve.~An example of simulated red-noise light curve is shown in Figure~\ref{sim-lc}.~We calculated corresponding power spectra searching for peaks in the 0.001--0.1~\rm{Hz} range.~The 3$\sigma$ confidence limits on the maximum power are calculated, including the uncertainties in the red noise model.~The chance probability of occurrence of the observed signal is obtained by counting the number of trial time series with powers equal to or exceeding the observed power in the frequency range 0.001--0.009~\rm{Hz}.~Our simulation results are shown in Figure~\ref{Monte-carlo}.~The detection significance for the observations showing the QPO is given in Table~\ref{peak}.~The detection significance for the observations showing the QPO is well above 3$\sigma$ for all observations except Obs~7,~Obs~13,~Obs~18 and \textsc{XRT} observation.


Figure~\ref{evol} shows the evolution of X-ray QPOs compared to optical QPOs observed during previous outbursts of this source.~Unlike optical QPOs, X-ray QPOs did not show a monotonic trend of dip frequency decreasing as the outburst declines. 


\begin{figure*}
\begin{minipage}{0.45\textwidth}
\includegraphics[height=3.in,width=8.cm,angle=0,keepaspectratio]{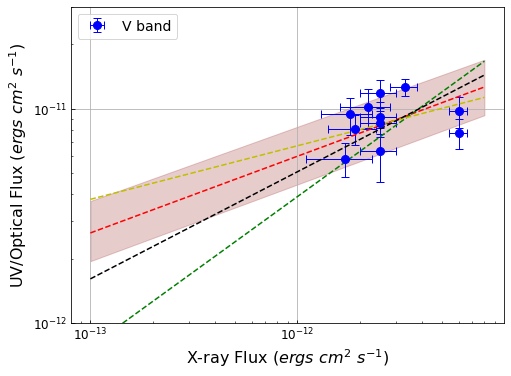}
\includegraphics[height=3.in,width=8.cm,angle=0,keepaspectratio]{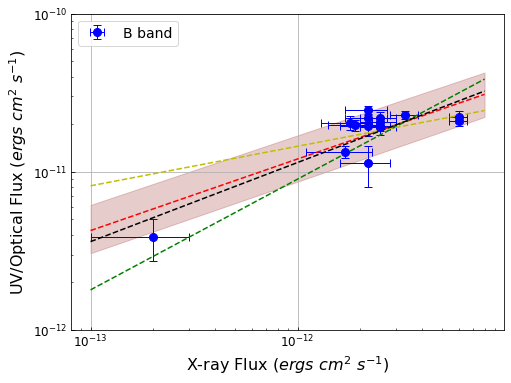}
\includegraphics[height=3.in,width=8.cm,angle=0,keepaspectratio]{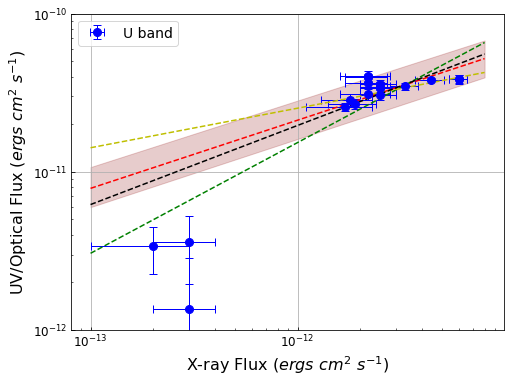}
\end{minipage}
\hspace{0.01\linewidth}
\begin{minipage}{0.45\textwidth}
\includegraphics[height=3.in,width=8.cm,angle=0,keepaspectratio]{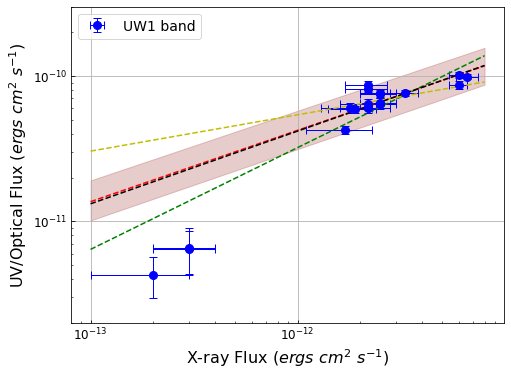}
\includegraphics[height=3.in,width=8.cm,angle=0,keepaspectratio]{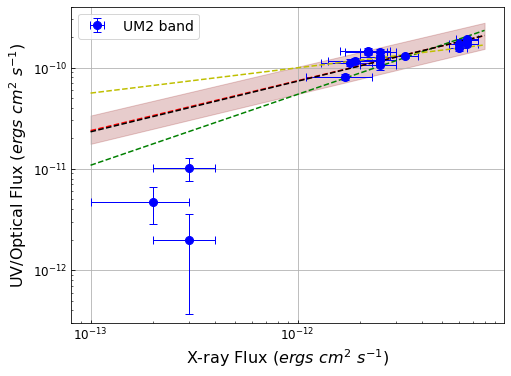}
\includegraphics[height=3.in,width=8.cm,angle=0,keepaspectratio]{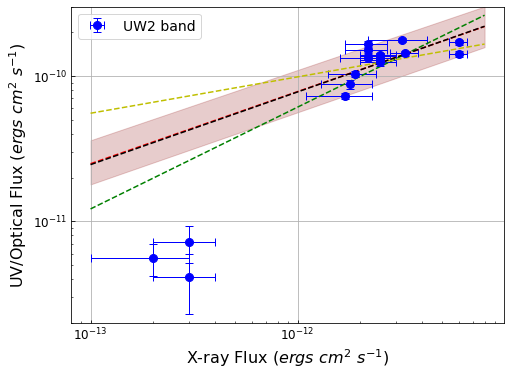}
\end{minipage}
\includegraphics[width=1.5\columnwidth]{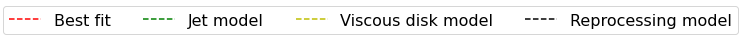}
\caption{Correlation slopes (${\beta}$) between the UVOT bands and the unabsorbed X-ray flux in the 0.5--10~keV energy band, with models for X-ray reprocessing, viscously heated disc and jet components.~The shaded region represents the 1-$\sigma$ confidence interval (refer to the text for more details).}
\label{Beta-models}
\end{figure*}

\begin{figure}
 \includegraphics[height=2\columnwidth,width=\columnwidth,angle=0,keepaspectratio]{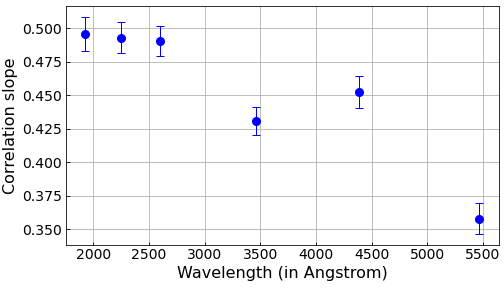}
 \caption{Best-fitting correlation slopes (${\beta}$) between the UVOT bands and the 0.5--10~\rm{keV} X-ray flux.}
 \label{correlation}
 \end{figure}
 
 \begin{table}
\caption{Peaks observed in the PDS shown in Figure~\ref{Periodogram} are listed below.}
\centering
\begin{tabular}{|c|c|c|}
\hline
Observation    &  Frequency~(mHz) & Significance~($\sigma$) \\
\hline
Obs~1     &   1.8$\pm$0.1     & 12  \\
Obs~2     &   1.2$\pm$0.1     & 8 \\
Obs~3     &   3.1$\pm$0.1     & 12 \\
Obs~4     &   5.4$\pm$0.1     & 3.9 \\
Obs~5     &   2.9$\pm$0.1     & 4.4  \\
Obs~6     &   1.6$\pm$0.1     & 12  \\
Obs~7     &   3.4$\pm$0.1     & 0.45 \\
Obs~8     &   2.7$\pm$0.1     & 14 \\
Obs~9     &   3.5$\pm$0.1     & 12 \\
Obs~10    &   1.9$\pm$0.1     & 5 \\
Obs~11    &   1.5$\pm$0.1     & 6 \\
Obs~12    &   3.7$\pm$0.1     & 7 \\
Obs~13    &   1.5$\pm$0.1     & 2.1 \\
Obs~14    &   4.1$\pm$0.1     & 7 \\
Obs~15    &   1.4$\pm$0.1     & 4.3 \\
Obs~16    &   3.5$\pm$0.1     & 6.7 \\
Obs~17    &   2.8$\pm$0.1     & 10 \\
Obs~18    &   2.5$\pm$0.1     & 0.4 \\
Obs~20    &   2.8$\pm$0.1     & 4.3 \\
XRT-31918084 &   1.8$\pm$0.1   & 2.7 \\

\hline
\end{tabular}
\label{peak}
\end{table}
 
\begin{table}
\caption{Correlation slope ($\beta$) between UV/optical and X-ray fluxes during 2019 outburst of J1357.}
\centering
\begin{tabular}{|c|c|c|}
\hline
Filter (wavelength; {\AA})     &   Best fit  slopes~($\beta$)  \\
\hline
V~(5468)     &   0.36 $\pm$ 0.01       \\
B~(4392)     &   0.45 $\pm$ 0.01       \\
U~(3465)     &   0.43 $\pm$ 0.01       \\
UVW1 (2600)     &   0.49 $\pm$ 0.01       \\
UVM2 (2246)     &   0.49 $\pm$ 0.01 \\
UVW2 (1928)     &   0.49 $\pm$ 0.01  \\
\hline
\end{tabular}
\label{Beta}
\end{table}

\section{UV/Optical and Radio Observations of J1357}
\label{sec:multi-band}
We also explored the multi-wavelength variability of J1357 during the 2019 outburst.~The bottom panel of Figure~\ref{PL-index} shows that the UV/optical magnitudes display a continuous decrease in brightness (from magnitude ${\sim}15$ to ${\sim}19$), similar to that observed in X-rays.~There is a correlation between X-ray and UV/optical fluxes~(Figure~\ref{Beta-models}).~We performed a Monte-Carlo simulation to compute the 1-$\sigma$ confidence interval (shaded region in Figure~\ref{Beta-models}).~10000 iterations were
performed and at each step, 1000 data points were
generated. We used the covariance matrix to determine confidence intervals of each random small subset of data, and this step was
repeated for 10000 iterations to obtain an average value of the confidence interval.~The best-fitting correlation slopes~($\beta$) are given in Table~\ref{Beta}.~We also observe that $\beta$ increases at shorter wavelengths, consistent with previous reports on this source (Figure~\ref{correlation}). 

We compared the correlation between the UV/optical and X-ray fluxes against correlations observed for three emission processes: X-ray reprocessing in the disc,~the viscously heated disc and jet emission.
For X-ray reprocessing, we adopt the theoretical model between the optical and X-ray luminosities given by \citet{Van94}.~According to this model
optical luminosity of an X-ray reprocessing accretion disc varies as 
$L_{\rm{opt}}$~${\propto}$~${L_{\rm{X}}}^{0.5}a$, where $a$ is the orbital separation of the system given by $3.5\times{10^{10}}(M_{\rm{BH}})^{1/3}(1+q)^{1/3}(P_{\rm{hr}})^{2/3}$ \citep{Frank02}.~The values of BH mass~($M_{\rm{BH}}$),~the mass ratio of the companion star to the compact object,~$q=M_C/M_{\rm{BH}}$ and the orbital period~($P_{\rm{hr}}$) were taken from \citet{Casares16}.~For the viscously heated disc and jet emission, we have used the following relations:~$L_{\rm{opt}}\propto~L_{\rm{X}}^{0.25}$ and $L_{\rm{opt}}\propto~L_{\rm{X}}^{0.7}$, respectively \citep[see][]{Russell06}.
The best-fitting correlation slopes (Table-\ref{Beta}) show that for all the \textsc{UVOT} bands the values lie closer to the model for X-ray reprocessing.~However, for the V band emission, there can also be a contribution from a viscously heated disc around a BH.

We did not detect the source at radio wavelengths with \emph{AMI}, our observations only yielding $3{\sigma}$ upper limits of 210 $\upmu$Jy beam$^{-1}$.
\section{Discussion}
\label{Discuss}
\subsection{Spectral Behaviour} 
\emph{Swift}-\textsc{XRT} and \emph{NICER} observations made during the 2019 outburst of J1357 were well-fit using a simple absorbed power-law consistent with the 2011 and 2017 outbursts of J1357 \citep{Armas13a, Beri19}.~We found that an additional disc component was needed to obtain the best fit in one of the \textsc{XRT} observations with a higher signal to noise ratio.~During the 2011 outburst of J1357, \citet{Armas13b} did find disc emission in their \emph{XMM-Newton} spectrum with a disc temperature of $\sim$ 0.2~$\mathrm{keV}$, similar to that measured here.~The low value of disc temperature observed during the outburst in 2011 was believed to be due to J1357 being in its low hard spectral state.~In fact, during the 2017 outburst, models fitted to \emph{NuSTAR} observations did not require any disc contribution.~The non-detection of the disc component was explained as being due to the presence of a cool disc \citep[see][for details]{Beri19}. 
Moreover, we did not find the presence of any reflection features in the form of an iron line in the \emph{NICER} spectra. This is again consistent with previous reports on this source \citep[see, e.g.][]{Beri19}. 


The exponential (viscous) decay timescale measured during earlier outbursts was found to be ${\sim}~64~\rm{d}$ \citep{Tetarenko2018}.~Moreover,~Figure~\ref{wijnands} shows that values of photon index are comparable to those observed during the decline of the 2011 and 2017 outbursts \citep{Armas13a, Beri19}.~During the 2019 outburst of J1357, our \emph{NICER} and \emph{Swift}-\textsc{XRT} observations revealed the maximum value of unabsorbed flux to be $\sim$ 8$\times$ $10^{-12}~\rm{erg~cm}^{-2}~\rm{s}^{-1}$ which corresponds to $L_{\rm X}$ of about $3.4{\times}10^{34}~($d$/6~{\rm kpc})^2{\rm erg}~{\rm s^{-1}}$.~This flux is almost 18 times lower than that observed in 2017 \citep{Beri19} and 55 times lower than measured in 2011 \citep{Armas13a}.~All this indicates that \emph{NICER} and \emph{Swift} observations were made during the declining phase of the outburst.

\subsection{Timing Behaviour}
Thanks to the excellent fast timing capability of \emph{NICER}, for the first time we were able to probe the X-ray variability in J1357.
Structures in the X-ray light curves were found to vary on timescales between ${\sim}200$ and $700~{\rm{s}}$ with a strong energy dependence. 

The \emph{NICER} PDS for Obs~9~(Figure~\ref{PDS}) showed the presence of a QPO
at around $3~\rm{mHz}$ with a $Q$ factor of $\sim$~4.~Early in the 2011 outburst of J1357, \emph{RXTE} found a similar QPO at $\sim$~6~$\mathrm{mHz}$ with $Q$~$\sim$~3 \citep{Armas13b}.~The presence of a sharp peak at corresponding frequencies was also observed in our Lomb-Scargle and CLEAN periodograms (see Figure~\ref{Periodogram}).~One of our most intriguing results is that we found peaks in the millihertz frequency range that were consistent with those seen in the optical during the decline phase of the 2011 and 2017 outbursts \citep{Corral13, Paice19}. 

It could be possible that the origin of X-ray QPOs is linked to that for optical dips, as they were also observed at similar frequencies during the 2019 outburst \citep[][]{Jimenez-Ibarra2019}.~During the previous outbursts of J1357, a monotonic decrease of the optical dip frequency as the outburst progresses has been observed; however, this behaviour is not evident in our X-ray data~(Figure~\ref{evol}).~The interpretation of this optical dip behaviour was that the inner edge of some obscuring material was moving outwards through the disc during the decline.~Thus, we cannot rule out other possibilities as to the origin of the X-ray QPO.~Millihertz X-ray QPOs observed in J1357 are believed to resemble `$1~\rm{Hz}$~QPOs'
seen in dipping neutron star systems \citep[see][]{Armas13b}, and it is therefore possible that they share a common origin.~Thus, millihertz X-ray QPOs observed in J1357 could be due to a structure in the inner disc which quasi-periodically obscures the inner region \citep{Jonker99} or due to relativistic Lense--Thirring precession of the inner accretion disc \citep{Homan12}.~For the first time millihertz X-ray QPOs have been significantly detected during the decline phase of the J1357 outburst, ruling out the earlier suggestion that the X-ray QPOs observed in J1357 are part of the new class of BH QPOs with frequencies in the millihertz range seen only at the start of an outburst \citep{Altamirano2012}.

\subsection{UV/Optical and X-ray correlation}
The decay profile after the outburst peak
is well explained using the disc-instability model including irradiation \citep{dubus1999,dubus2001} and such profiles have been observed in a number of 
BH-LMXB outburst light-curves, including
outburst light-curves of J1357 \citep{Tetarenko2018}. Therefore, as reported in \citet{Beri19}, we compared the correlation between the UV/optical and X-ray fluxes against correlations observed for three emission processes: X-ray reprocessing in the disc,~the viscously heated disc, and jet emission~(Figure~\ref{Beta-models}).~For the V band, the best-fit {$\beta$} value was found to
deviate from the predicted values for the reprocessing and
jet models~(Figure~\ref{correlation}).~It might be possible that intrinsic thermal emission from the viscously heated outer accretion disc contributes significant light in the optical \citep{Frank02}.
However, for the other bands the $\beta$ values were more consistent with the predicted values for the reprocessing model, favouring the idea suggested by \citet{Van94} that for smaller accretion discs (i.e. smaller P$_{\rm orb}$) a higher value of the average surface temperature of the
disc is expected (presumably as it is closer to the compact object and irradiating source). Therefore, one would expect to find a larger fraction of the reprocessed emission in the UV band. 

\section{Conclusions}
In this work, we have analysed multiwavelength data of J1357 during its 2019 outburst.~We summarise our findings as follows:
\begin{itemize}
\item Our X-ray spectral analysis suggests that our \emph{NICER} and \emph{Swift}-\textsc{XRT} observations were made during the declining phase of the outburst.~The maximum X-ray flux we observed is almost 18 times lower than the peak seen in 2017 \citep{Beri19} and 55 times lower than that in 2011 \citep{Armas13a}.~Moreover, our spectral index measurements are comparable to those observed during the decay phase observations in 2011 and 2017.~One of the \textsc{XRT} observations also revealed the presence of a cool accretion disc. \\
\item X-ray QPOs in the millihertz frequency range have been detected for the first time throughout the outburst of J1357.~This is in contrast to that observed in the 2011 outburst, during which $6~\rm{mHz}$ QPO was detected only during the first \emph{RXTE} observation and was not present during any of the later RXTE observations or during the
XMM-Newton observation which was taken 3 days after the first RXTE observation \citep{Armas13b}.~A number of models could explain this X-ray variability, not least of which involves an X-ray component to the curious dips that, so far, have chiefly been seen only in the
optical. Therefore, future X-ray observations of such outbursts would help to answer these questions. \\
\item Our optical/UV and X-ray correlation study indicates a significant contribution of the X-ray reprocessing to the optical and UV emission.~However, this is in contrast to what has been observed during the 2011 and 2017 outbursts of J1357 where \citet{Armas13a} and \citet{Beri19} observed UV/optical emission to be dominated by the viscously
heated disc, with little or no contribution from X-ray irradiation. 
\end{itemize}

\section*{Acknowledgements}
We would like to thank the anonymous referee for their thorough comments that helped us in improving the quality of the paper.
 AB is grateful to the Royal Society, United Kingdom. She is supported by an INSPIRE Faculty grant (DST/INSPIRE/04/2018/001265) by the Department of Science and Technology, Govt. of India.~She would also like to thank Dr. Jakob van den Eijnden for fruitful discussions. This work is based on data from the \emph{NICER} and \emph{Swift} mission. We would like to thank all the members of \emph{NICER} and \emph{Swift} team for TOO observations. This research has made use of data and/or software provided by the High Energy Astrophysics Science Archive Research Center (HEASARC), which is a service of the Astrophysics Science
Division at NASA/GSFC and the High Energy Astrophysics Division of the Smithsonian Astrophysical Observatory.~This research has made use of NASA's Astrophysics Data System.~We thank the staff of the Mullard Radio Astronomy Observatory for their assistance in the commissioning, maintenance and operation of AMI, which is supported by the Universities of Cambridge and Oxford.
We also acknowledge support from the European Research Council under grant ERC-2012-StG-307215 LODESTONE. \\
This work made use of the NumPy \citep{Vanderwalt2011}, Matplotlib \citep{Hunter2007}, Scipy \citep{jones_scipy2001}, pandas \citep{mckinney-proc-scipy-2010}, Astropy \citep{Astropy2013}, Stingray \citep{Huppenkothen2019b} Python packages.

\section*{DATA AVAILABILITY}
The data underlying this article
are publicly available from  the  High  Energy  Astrophysics  Science Archive Research Center (HEASARC),~provided by NASA's Goddard Space Flight Center.

\bibliography{ref}{}
\bibliographystyle{mnras}
\nopagebreak
\appendix \section{Light curve Simulation Results}

\begin{figure*}
\begin{minipage}{0.45\textwidth}
\includegraphics[height=\linewidth,width=\linewidth,angle=0,keepaspectratio]{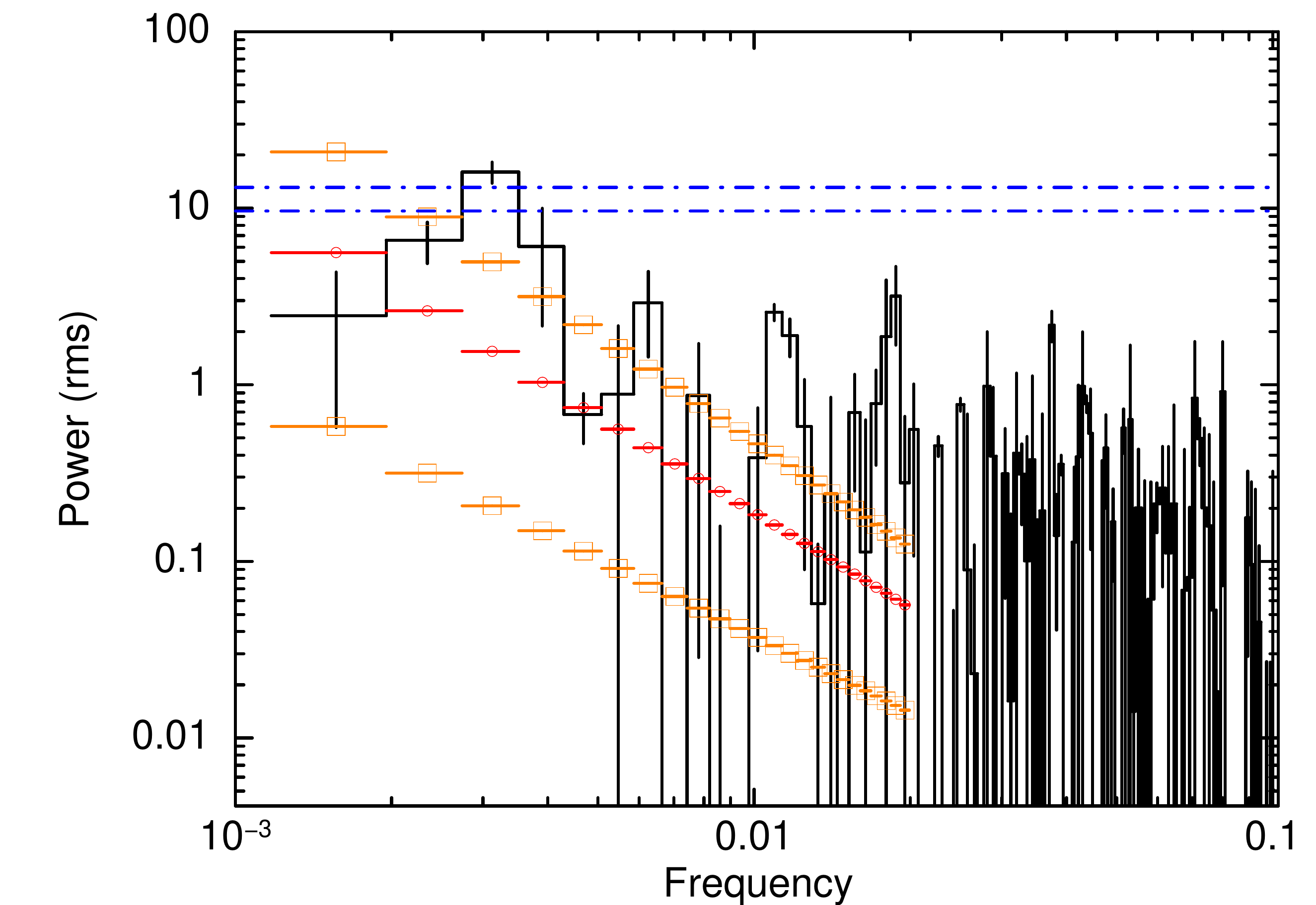}
\end{minipage}
\hspace{0.01\linewidth}
\begin{minipage}{0.45\textwidth}
\includegraphics[height=\linewidth,width=\linewidth,angle=0,keepaspectratio]{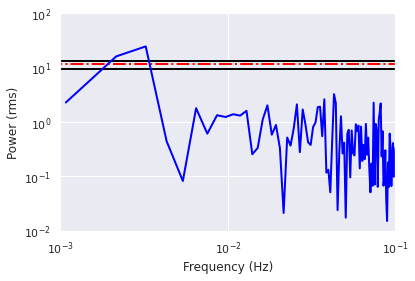}
\end{minipage}
\caption{The left plot shows the PDS created using \textit{`powspec'} ftool (as shown in Figure~\ref{PDS}), orange squares are the uncertainties in the red noise model while the best fitting photon index is shown in red circles.
These values were used to simulate light curves, and blue dashed lines is the maximum power observed at 3~$\sigma$ confidence using the observed range of photon index.~On the right, we show an example PDS created using light curve of Obs~9, using stingray library \textit{`powerspectrum'}. In order to compute significance of observed QPO peak, we have generated PDS of simulated light curves using this tool.~The dotted dashed lines in red indicate the maximum power observed at 3~$\sigma$ confidence in the simulated light curves, using the best-fitting photon index and solid black lines are calculated, including the uncertainties in the red noise model.}
\label{Obs109-sig}
\end{figure*}

\begin{figure*}
\begin{minipage}{0.45\textwidth}
\includegraphics[height=4in,width=3in,angle=0,keepaspectratio]{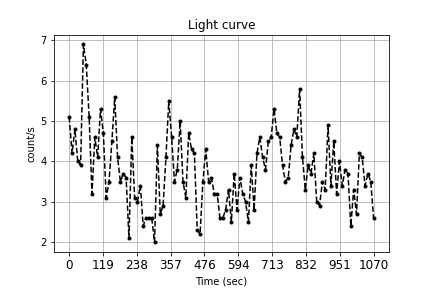}
\end{minipage}
\hspace{0.01\linewidth}
\begin{minipage}{0.45\textwidth}
\includegraphics[height=4in,width=3in,angle=0,keepaspectratio]{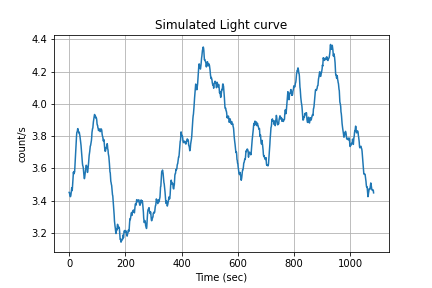}
\end{minipage}
\caption{The left plot is the light curve of one of the observations~(Obs~8) while the right plot is the simulated red-noise light curve obtained using the best-fitting parameters of the observed light curve.}
\label{sim-lc}
\end{figure*}

\begin{figure*}
\begin{minipage}{0.45\textwidth}
\includegraphics[height=5in,width=3in,angle=0,keepaspectratio]{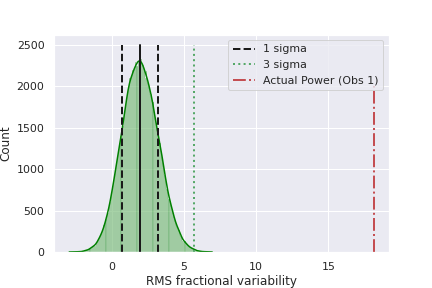}
\includegraphics[height=5in,width=3in,angle=0,keepaspectratio]{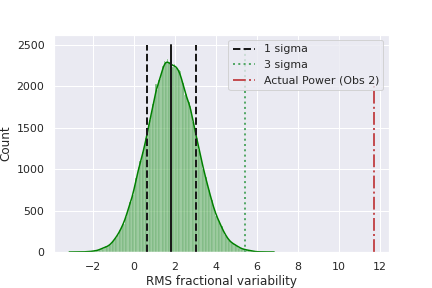}
\includegraphics[height=5in,width=3in,angle=0,keepaspectratio]{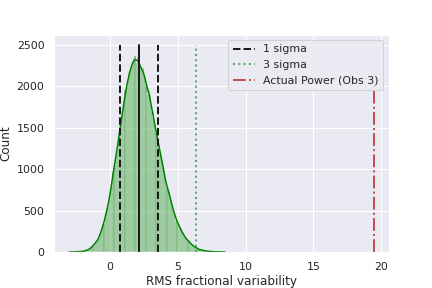}
\includegraphics[height=5in,width=3in,angle=0,keepaspectratio]{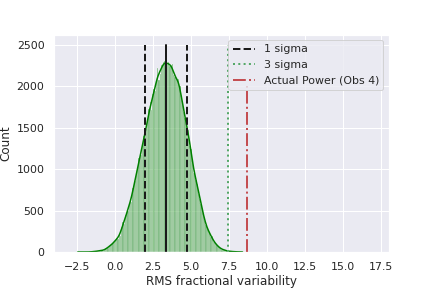}
\end{minipage}
\hspace{0.01\linewidth}
\begin{minipage}{0.45\textwidth}
\includegraphics[height=5in,width=3in,angle=0,keepaspectratio]{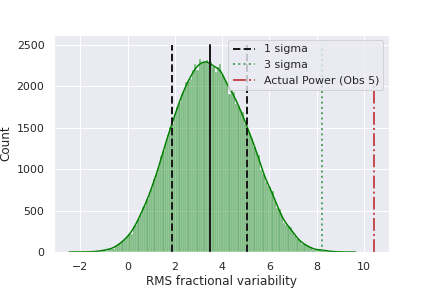}
\includegraphics[height=5in,width=3in,angle=0,keepaspectratio]{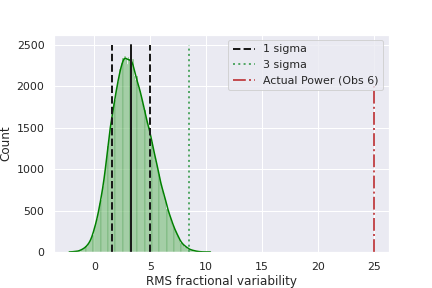}
\includegraphics[height=5in,width=3in,angle=0,keepaspectratio]{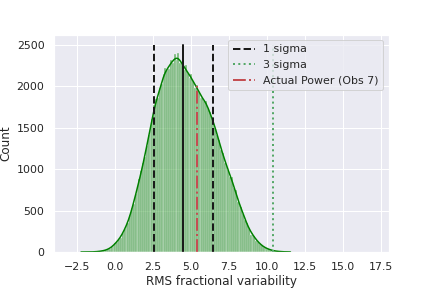}
\includegraphics[height=5in,width=3in,angle=0,keepaspectratio]{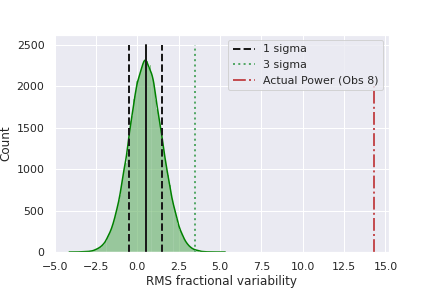}
\end{minipage}
\end{figure*}

\begin{figure*}
\begin{minipage}{0.45\textwidth}
\includegraphics[height=5in,width=3in,angle=0,keepaspectratio]{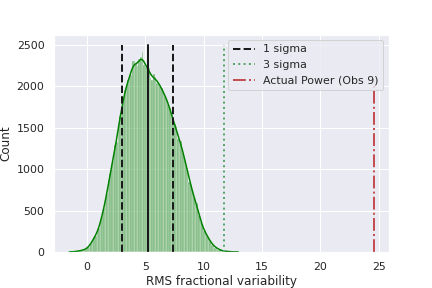}
\includegraphics[height=5in,width=3in,angle=0,keepaspectratio]{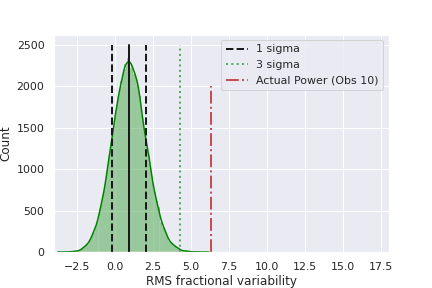}
\includegraphics[height=5in,width=3in,angle=0,keepaspectratio]{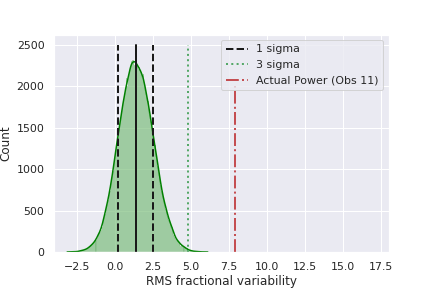}
\includegraphics[height=5in,width=3in,angle=0,keepaspectratio]{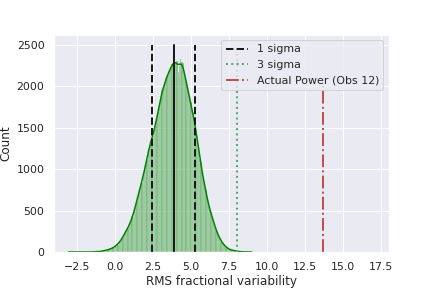}
\end{minipage}
\hspace{0.01\linewidth}
\begin{minipage}{0.45\textwidth}
\includegraphics[height=5in,width=3in,angle=0,keepaspectratio]{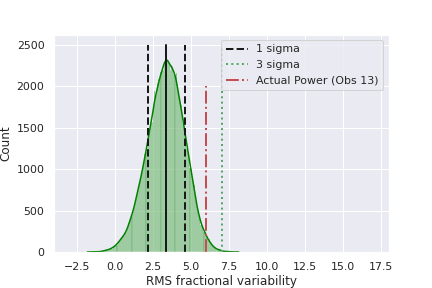}
\includegraphics[height=5in,width=3in,angle=0,keepaspectratio]{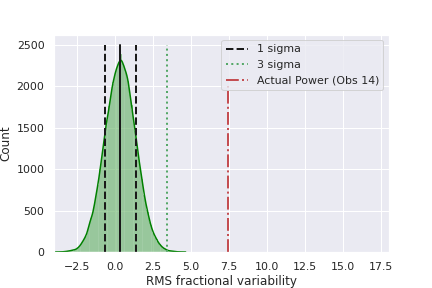}
\includegraphics[height=5in,width=3in,angle=0,keepaspectratio]{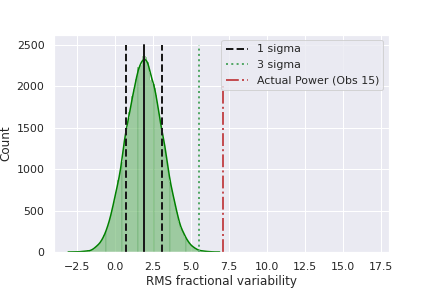}
\includegraphics[height=5in,width=3in,angle=0,keepaspectratio]{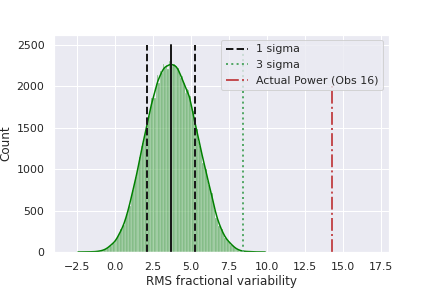}
\end{minipage}
\end{figure*}

\begin{figure*}
\begin{minipage}{0.45\textwidth}
\includegraphics[height=5in,width=3in,angle=0,keepaspectratio]{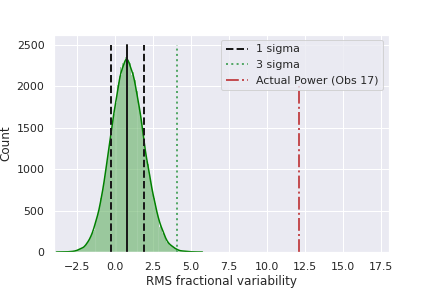}
\includegraphics[height=5in,width=3in,angle=0,keepaspectratio]{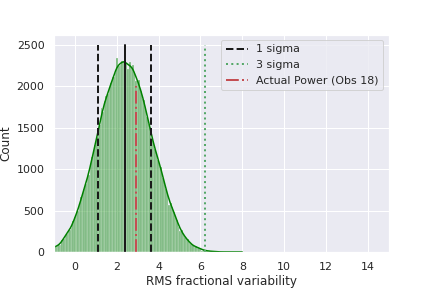}
\end{minipage}
\hspace{0.01\linewidth}
\begin{minipage}{0.45\textwidth}
\includegraphics[height=5in,width=3in,angle=0,keepaspectratio]{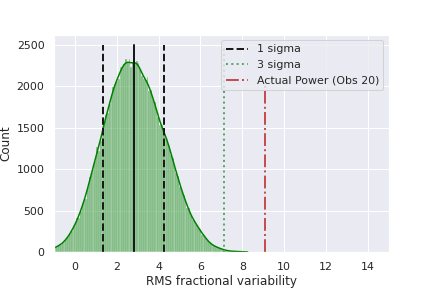}
\includegraphics[height=5in,width=3in,angle=0,keepaspectratio]{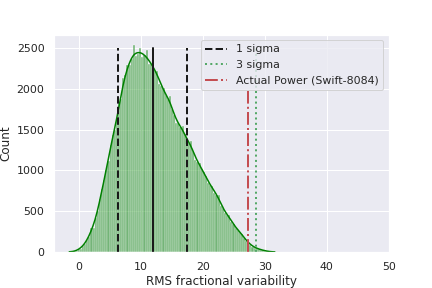}
\end{minipage}
\caption{Results from Monte-Carlo simulations for testing significance of the QPO peak in the periodogram.~For each simulation, we compared power in the frequency range 0.001--0.009~\rm{Hz} to that observed in the PDS of real light curve.~We have used \textit{Powerspectrum} of Stingray package to create power density spectra normalized to squared fractional rms.~The red dashed lines represent actual power observed in real light curves.}
\label{Monte-carlo}
\end{figure*}

\label{lastpage}
\end{document}